\documentclass[twocolumn]{aastex61}

\newcommand\kms{\hbox{km$\,$s$^{-1}$}}
\newcommand\masyr{\hbox{mas$\ $yr$^{-1}$}}

\received{March 7, 2017}
\revised{March 17, 2017}
\accepted{March 17, 2017}

\submitjournal{ApJ}

\shorttitle{Distance to the Low-velocity Cloud}
\shortauthors{Iwata et al.}

\begin{document}

\title{Distance to the Low-velocity Cloud in the Direction of the High-velocity Compact Cloud CO--0.40--0.22}

\correspondingauthor{Yuhei Iwata}

\email{iwa-astro4338@z3.keio.jp}

\author[0000-0002-9255-4742]{Yuhei Iwata}
\affil{School of Fundamental Science and Technology, Graduate School of Science and
Technology, Keio University, 3-14-1 Hiyoshi, Kohoku-ku, Yokohama, Kanagawa 223-8522, Japan}

\author{Haruka Kato}
\affiliation{Department of Physics, Faculty of Science and Technology, Keio University, 3-14-1
Hiyoshi, Kohoku-ku, Yokohama, Kanagawa 223-8522, Japan}

\author{Daisuke Sakai}
\affiliation{Department of Astronomy, Graduate School of Science, The University of Tokyo, 7-3-1 Hongo, Bunkyo-ku, Tokyo 113-0033, Japan}
\affiliation{Mizusawa VLBI Observatory, National Astronomical Observatory of Japan, 2-12 Hoshi-ga-oka, Mizusawa-ku, Oshu-shi, Iwate 023-0861, Japan}

\author{Tomoharu Oka}
\affiliation{School of Fundamental Science and Technology, Graduate School of Science and
Technology, Keio University, 3-14-1 Hiyoshi, Kohoku-ku, Yokohama, Kanagawa 223-8522, Japan}
\affiliation{Department of Physics, Faculty of Science and Technology, Keio University, 3-14-1
Hiyoshi, Kohoku-ku, Yokohama, Kanagawa 223-8522, Japan}

\begin{abstract}
CO--0.40--0.22 is a peculiar molecular cloud that is compact and has an extraordinary broad velocity width. It is found in the central molecular zone of our Galaxy.  In this direction, there is another cloud with an H$_2$O maser spot at a lower velocity.  Collision with this low-velocity cloud could be responsible for the broad velocity width of CO--0.40--0.22.  We performed phase-referencing VLBI astrometry with VERA and detected the annual parallax of the H$_2$O maser spot in the low-velocity cloud to be $0.33 \pm 0.14$ mas, which corresponds to a distance of $3.07^{+2.22}_{-0.91}$ kpc from the Sun.  This implies that the low-velocity cloud is located in the Galactic disk on the near side of the central molecular zone.
\end{abstract}
\keywords{astrometry --- Galaxy: center --- ISM: clouds --- masers}

\section{Introduction} \label{sec:intro}
Large-scale CO surveys of the central molecular zone (CMZ) of our Galaxy have detected a peculiar population of molecular clouds, namely, high-velocity compact clouds \citep[HVCCs;][]{Oka98,Oka07,Oka12}.  HVCCs are characterized by their compact appearance ($d\!<\!10$ pc) and unusually large velocity width ($\Delta V\!>\!50$ \kms).  Some of energetic HVCCs contain small expanding arcs and shells, indicating that local explosive events, such as supernova explosions, may be responsible for the origin of HVCCs \citep{Oka99, Oka01, Oka08, Tanaka07}.  

\begin{figure}[ht!]
\plotone{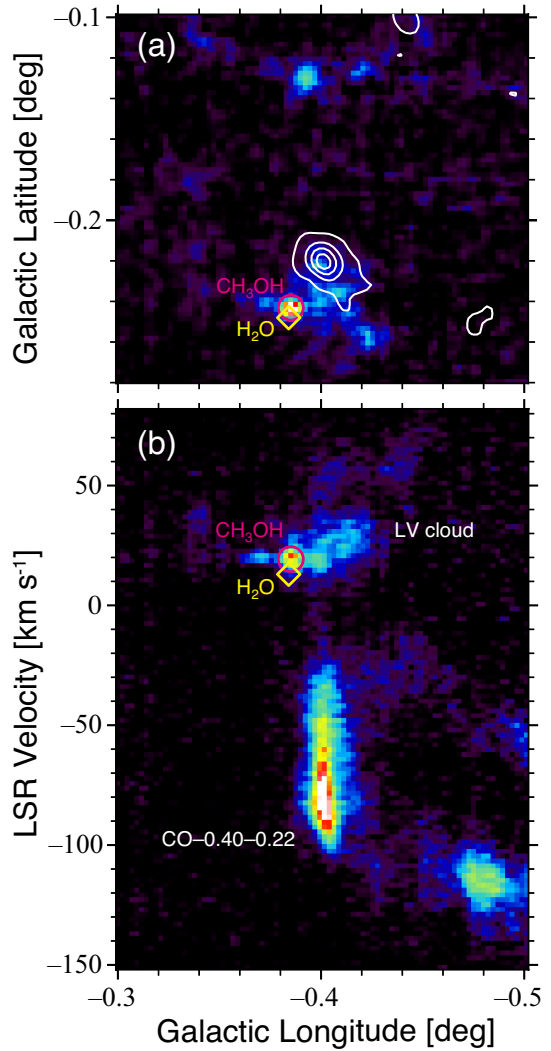}
\caption{(a) Map of velocity-integrated HCN {\it J}=4--3 emission.  The velocity ranges are taken from $V_{\rm LSR}\!=\!0$ to $+40$ \kms (color) and $V_{\rm LSR}\!=\!-120$ to 0 \kms (contour).  The intensity increases as the color changes from violet to white.  The contour levels are 100, 300, 500, and 700 K\ \kms.  The yellow diamond denotes the position of the 22 GHz H$_2$O maser spot at H$_2$O 359.62--0.25 \citep{Caswell83}, while the magenta circle denotes that of the 6.7 GHz CH$_3$OH maser spot \citep{Chambers14}.
(b) Longitude-velocity map of HCN {\it J}=4--3 emission integrated over latitudes from $-0\fdg28$ to $-0\fdg20$.  Yellow diamond and magenta circle denote the longitude and velocity of H$_2$O and CH$_3$OH maser spots, respectively.
\label{fig:1}}
\end{figure}

However, a majority of HVCCs do not have any expanding features or counterparts at other wavelengths.  CO--0.40--0.22 is such a featureless, energetic HVCC centered at $(l, b, V_{\rm LSR})\!=\!(-0\fdg40, -0\fdg22, -80\,\kms)$.  It has a small size ($\sim\!3$ pc; Figure \ref{fig:1}a) and an extremely broad velocity width ($\sim\!100$ \kms; Figure \ref{fig:1}b).  Intensive studies based on single-dish observations of molecular lines revealed that CO--0.40--0.22 consists of an intense component with a shallow velocity gradient and a less intense high-velocity wing.  This behavior is interpreted as a gravitational kick to the molecular cloud caused by an invisible compact object with a mass of $\sim\!10^5 \,M_{\odot}$ \citep{Oka16}.  This massive object could be an intermediate-mass black hole (IMBH), which would contribute to the evolution of the central supermassive black hole \citep[SMBH;][]{Ebisuzaki01}.

In the direction of CO--0.40--0.22, there is another molecular cloud at $V_{\rm LSR}\!\sim\!+20$ \kms\ (Figure \ref{fig:1}b).  This low-velocity cloud (LV cloud) is associated with a bright nebula of infrared emission, 22 GHz H$_2$O maser emission, and 6.7 GHz CH$_3$OH maser emission \citep{Caswell83, Chambers14}, indicating that massive star formation is taking place. Its physical relation to CO--0.40--0.22 is unclear.  If the LV cloud is physically associated with CO--0.40--0.22, collision between these clouds with significantly different velocities induces intense shock, resulting in a broad velocity width feature, which bridges colliding clouds in the position-velocity space \citep[e.g.,][]{Matsumura12}.  To examine the cloud-cloud collision scenario for CO--0.40--0.22, the distance to the LV cloud should be determined directly.  

\section{Observations and analysis} \label{sec:obs}
Observations of H$_2$O 6$_{16}$--5$_{23}$ (22.235080 GHz) maser in H$_2$O 359.62--0.25 were carried out in 12 epochs from November 2015 to January 2017 with VERA \footnote{VLBI Exploration of Radio Astrometry (VERA) is a Japanese VLBI array for phase-referencing astrometry, consisting of four domestic stations equipped with dual-beamed radio telescopes.}.  Observational details are summarized in Table 1.  The interval of the observations was $\sim$ 1 month.  All four stations of VERA were utilized except for epochs 4 and 5.  H$_2$O 359.62--0.25 and position reference source J1745--2820 $(\alpha_{\rm J2000.0}\!=\!17^{\rm h}45^{\rm m}52\fs4968, \delta_{\rm J2000.0}\!=\!-28\arcdeg20\arcmin26\farcs294)$ were observed simultaneously in the dual-beam mode.  The tracking center position of H$_2$O 359.62--0.25 was set to $(17^{\rm h}45^{\rm m}39\fs0908, -28\arcdeg23\arcmin30\farcs218)$.
The separation angle between them was 1\fdg05.  NRAO530 was also observed as a bandpass and delay calibrator every 40 min.  The observation sequence in each day spanned $\sim$ 3 h.  

\begin{deluxetable}{cccc}
\tablecaption{VERA Observations. \label{tab:1}}
\tablehead{
\colhead{Epoch} & \colhead{Code} & \colhead{Date} & \colhead{Antennas Available\tablenotemark{a}}
}
\startdata
1 & r15332a & 2015 Nov 28 & MZ, IR, OG, IS \\
2 & r15360a & 2015 Dec 26 & MZ, IR, OG, IS \\
3 & r16026b & 2016 Jan 26 & MZ, IR, OG, IS \\
4 & r16061e & 2016 Mar 1 & IR, OG, IS \\
5 & r16084c & 2016 Mar 24 & IR, OG, IS \\
6 & r16114b & 2016 Apr 23 & MZ, IR, OG, IS \\
7 & r16142b & 2016 May 21 & MZ, IR, OG, IS\\
8 & r16166a & 2016 Jun 14 & MZ, IR, OG, IS\\
9 & r16236a & 2016 Aug 23  & MZ, IR, OG, IS\\
10 & r16262a & 2016 Sep 18 & MZ, IR, OG, IS \\
11 & r16327a & 2016 Nov 22 & MZ, IR, OG, IS \\
12 & r17017a & 2017 Jan 17  & MZ, IR, OG, IS \\
\enddata
\tablenotetext{a}{Antenna codes are MZ: at Mizusawa, Iwate, IR: at Iriki, Kagoshima, OG: at Ogasawara Islands, Tokyo, and IS: at Ishigakijima, Okinawa.}
\end{deluxetable}

Left-handed circular polarization was received, sampled with 2-bit quantization.  The signals were filtered using the VERA digital filter unit \citep{Iguchi05} to obtain 16 IF channels each with a bandwidth of 16 MHz.  One IF channel was assigned to H$_2$O 359.62--0.25 and 15 IF channels to J1745--2820.  The data were recorded onto magnetic tapes at a rate of 1024 Mbps.  The system temperatures including atmospheric loss were measured to be 100--1000 K, depending on the stations and weather conditions.  Correlation processing was conducted on the Mizusawa software correlator at the National Astronomical Observatory of Japan (NAOJ).  The correlator accumulation period was 1 s.  For the J1745--2820 data, the spectral resolution was 64 points per each 16 MHz channel.  For the H$_2$O 359.62--0.25 data, the frequency resolution was 15.625 kHz, which corresponds to a velocity resolution of 0.21 \kms.  

Data reductions were performed by using the NRAO Astronomical Image Processing System (AIPS).  The amplitude was calibrated by the system noise temperatures, and bandpass calibration was made with the NRAO530 data.  The dual-beam phase-calibration data and the modified delay-tracking model were applied for accurate measurements.  For phase-referencing, we calibrated the clock parameters using NRAO530 and performed a fringe fitting on the maser spot in H$_2$O 359.62--0.25 because of the low intensity of J1745--2820 ($\sim$ 70 mJy).  The solutions were applied to the J1745--2820 data.  We obtained the image of J1745--2820 by using task IMAGR.  The position of J1745--2820 with respect to the H$_2$O maser spot was determined by elliptical Gaussian fitting to the brightness peak of the image.  Since the position of J1745--2820 should be stable, we can regard the positional change of J1745--2820 as that of the H$_2$O maser spot.  

\section{Results} \label{seq:result}
\begin{figure}
\plotone{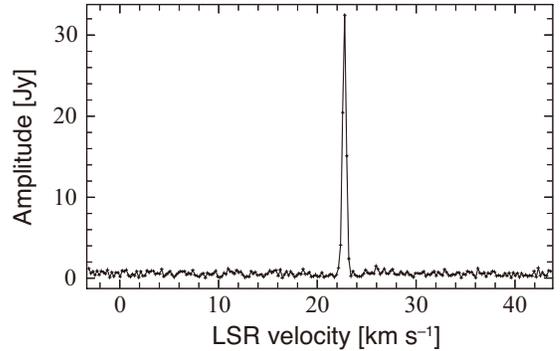}
\caption{The cross-power spectrum of H$_2$O 359.62--0.25.  This was obtained from the data on the sixth epoch at Mizusawa-Iriki baseline.
\label{fig:2}}
\end{figure}

\begin{figure}
\plotone{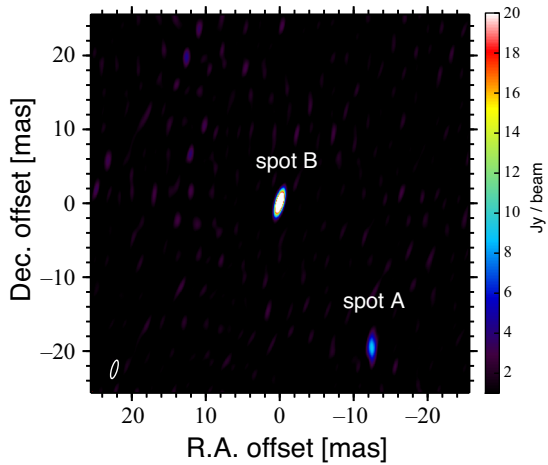}
\caption{Phase-referenced image of the H$_2$O maser spots of $V_{\rm LSR}\!=\!22.7$ \kms\ at the sixth epoch.  Maser spot A was visible during epochs 1--8, while spot B was visible during epochs 6--8.  The coordinate offsets are with respect to the position of spot B ($\alpha_{\rm J2000.0}\!=\!17^{\rm h}45^{\rm m}39\fs0918, \delta_{\rm J2000.0}\!=\!-29\arcdeg23\arcmin30\farcs207$).  The synthesized beam is shown in the bottom left corner.  
\label{fig:3}}
\end{figure}

H$_2$O 6$_{16}$--5$_{23}$ maser line was detected from H$_2$O 359.62--0.25 at $V_{\rm LSR}\!\sim\!+22.7$ \kms\ in the first eight epochs (Figure \ref{fig:2}).  Only one spot (named `A') was visible during epochs 1--5, while another spot (named `B') appears at $\sim 20$ \added{mas} northeast of spot A during epochs 6--8 (Figure \ref{fig:3}).  We traced the position of spot A with respect to J1745--2820. Since spot B was more intense than A during epochs 6--8, we traced it with respect to J1745--2820.  This was because we used the H$_2$O 359.62--0.25 data for the fringe fitting process.  For this reason, we added the positional offset between spots A and B to obtain the accurate position of spot A.

\begin{figure}
\plotone{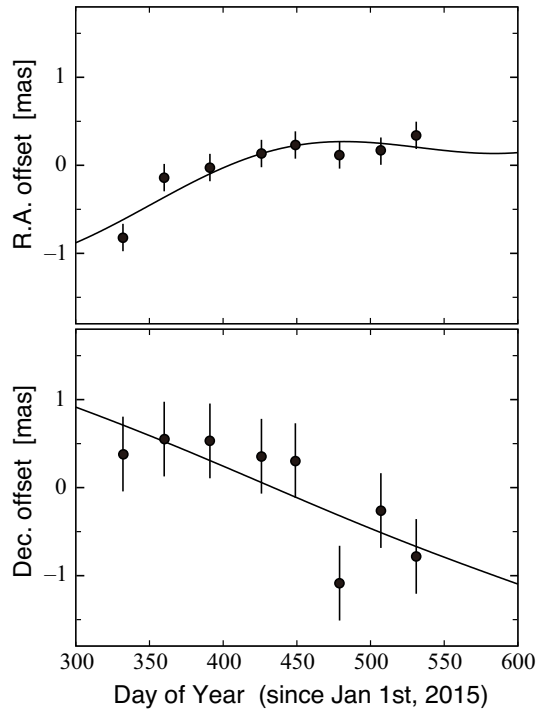}
\caption{
Positional variations of H$_2$O maser spot A.  The top figure is for the R.A. direction and the bottom figure is for the Dec. direction.  The solid lines are the best-fitting models including the annual parallax and the linear proper motion.  Errors are the standard deviations from the best-fit curves.  The position offsets are with respect to the average position of epochs 1--8 $(\alpha_{\rm J2000.0}\!=\!17^{\rm h}45^{\rm m}39\fs0908, \delta_{\rm J2000.0}\!=\!-29\arcdeg23\arcmin30\farcs226$).
\label{fig:4}}
\end{figure}

Figure \ref{fig:4} shows the results of the position measurements of maser spot A.  The position offsets are with respect to the average position of epochs 1--8.  As shown in Figure \ref{fig:4}, we detected the movement of spot A, particularly in the R.A. plot, which certainly deviates from a linear motion.  The deviation seems to be a sinusoidal, annual modulation.  This must be due to the annual parallax of the maser spot.  We derived the annual parallax and proper motions using both the R.A. and Dec. data simultaneously.  Table \ref{tab:2} summarizes the best-fitting result.  The obtained annual parallax and proper motions are $0.33 \pm 0.14$ mas and $1.31 \pm 0.33$ \masyr\ for R.A. and $-2.41 \pm 0.87$ \masyr\ for Dec., respectively.  We employed a uniform weighting in the fitting procedure, because systematic errors are considerably larger than the statistical errors in the VLBI observations.  We evaluated the systematic errors to be 0.16 mas in R.A. and 0.43 mas in Dec.  A worse astrometric accuracy in Dec. can be seen in low-elevation angle sources, such as Galactic center sources, caused by an atmospheric zenith delay residual \citep{Honma08}.  The derived parallax corresponds to a distance of $3.07^{+2.22}_{-0.91}$ kpc from the Sun.  

\begin{deluxetable*}{cccccc}
\tablecaption{The best-fitting values of Parallax $\pi$ and proper motions $\mu_\alpha \cos{\delta}$ and $\mu_\delta$. \label{tab:2}}
\tablehead{
\colhead{$V_{\rm LSR}$ (\kms)} & \colhead{$\pi$ (mas)} & \colhead{Distance (kpc)} & \colhead{$\mu_{\alpha}\cos{\delta}$ (\masyr)} & \colhead{$\mu_{\delta}$ (\masyr)}
}
\startdata
+22.7 & $0.33 \pm 0.14$ & $3.07^{+2.22}_{-0.91}$ & $1.31 \pm 0.33$ & $-2.41 \pm 0.87$ \\
\enddata
\end{deluxetable*}

\section{Discussion and Summary}\label{seq:discussion}
The derived distance from the Sun corresponds to a Galactocentric distance of $5.26^{+0.91}_{-2.22}$ kpc if we assume $R_{0}\!=\!8.33$ kpc \citep{Gillessen09}.  This shows that H$_2$O 359.62--0.25, as well as the LV cloud, is not in the CMZ. Assuming the normal distribution of the observed parallax, we obtained the probability that H$_2$O 359.62--0.25 is located further than 8 kpc to be 6.3\%.  Thus, the LV cloud is not in the CMZ with a confidence level of greater than 93.7\%.  

\begin{figure}
\plotone{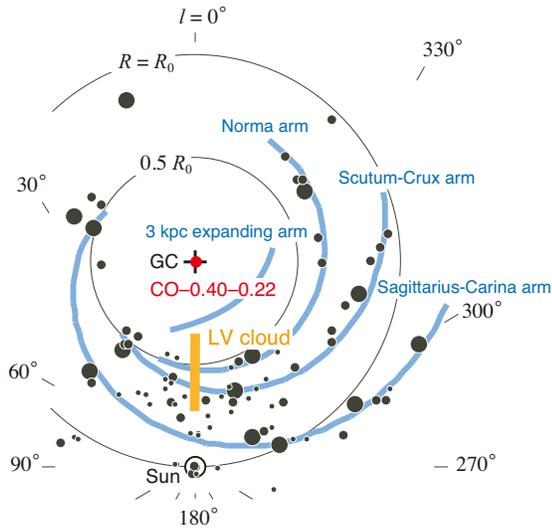}
\caption{Locations of CO--0.40--0.22 and the LV cloud on the Galactic plane.  Orange solid line indicates the derived distance of the LV cloud with $1\sigma$ uncertainty.  Blue solid lines indicate the spiral arms of our Galaxy.  The shapes of the spiral arms except for the 3 kpc expanding arm were obtained by fitting logarithmic spirals to high-excitation parameter H{\sc ii} regions.  Black filled circles indicate the loci of the H{\sc ii} regions, with their sizes representing the excitation classification as described in \citet{Georgelin76}.  The shape of the 3 kpc expanding arm was obtained from an artist's conception of the Milky Way (R. Hurt: NASA/JPL-Caltech/SSC).
\label{fig:5}}
\end{figure}

Figure \ref{fig:5} shows the line-of-sight location of the LV cloud.  The most likelihood value $R_{\rm G}\!=\!5.92$ kpc falls on the Scutum-Crux arm.  On the other hand, the velocity of the LV cloud ($V_{\rm LSR} \sim +20$ \kms) implies that the LV cloud may belong to the 4 kpc molecular ring \citep{Sofue06}, which is also within the $1\sigma$ uncertainty range.  The LV cloud could be a member of the OB association responsible for RCW 137, which is a large (diameter $\sim 0\fdg3$) H{\sc ii} region centered at $(l, b)\!=\!(-0\fdg2, -0\fdg2)$ \citep{Rodgers60}.  RCW 137 is visible in the optical wavelength, indicating a distance of $\lesssim 3$ kpc from the Sun.  All these facts support the near distance of the LV cloud.  

Our parallax measurement of H$_2$O 359.62--0.25 concludes that the LV cloud is in the Galactic disk on the near side of the CMZ.  On the other hand, CO--0.40--0.22 is supposed to be in the CMZ by its extraordinary broad velocity width.  An absorption feature in CO lines at $V_{\rm LSR} \sim -50$ \kms\ \citep{Oka12} indicates that CO--0.40--0.22 is certainly behind the 3 kpc expanding arm \citep{Sofue06}.  Therefore, the LV cloud is irrelevant to CO--0.40--0.22.  The cloud-cloud collision scenario seems to be not applicable to the CO--0.40--0.22 case.  We supposed another formation scenario for CO--0.40--0.22, a gravitational kick by a massive, point-like object, which could be an IMBH \citep{Oka16}.  In summary, this annual parallax study \replaced{gave a vote to the gravitational kick scenario for CO--0.40--0.22.}{ruled out at least one formation scenario other than the gravitational kick scenario.}

\acknowledgments
We are grateful to all VERA staff members for operation of the telescope and the data correlation.  The data analysis was in part carried out on the open use data analysis computer system at the Astronomy Data Center (ADC) of the NAOJ.  We thank the anonymous referee for helpful comments.  T.O. acknowledges support from JSPS Grant-in-Aid for Scientific Research (B) No. 15H03643.

\end{document}